\documentclass[pss]{wiley2sp} 
\usepackage{bm}

\begin{document}

\title{Pressure-induced structural transitions in multi-walled carbon nanotubes}
\titlerunning{Pressure-induced structural transitions in multi-walled carbon nanotubes}

\author{
  Hiroyuki Shima\textsuperscript{\textsf{\bfseries 1,\Ast}} and
  Motohiro Sato\textsuperscript{\textsf{\bfseries 2}}}

\authorrunning{H. Shima and M. Sato}

\mail{e-mail
  \textsf{shima@eng.hokudai.ac.jp}, Phone
  +81-11-706-6624, Fax +81-11-706-6859}

\institute{%
\textsuperscript{1}\,
Department of Applied Physics, Graduate School of Engineering,
Hokkaido University, 060-8628 Japan\\
\textsuperscript{2}\,
Department of Socio-Environmental Engineering, Graduate School of Engineering, 
Hokkaido University, Sapporo 060-8628, Japan}

\received{XXXX, revised XXXX, accepted XXXX} 
\published{XXXX} 


\abstract{
\abstcol{
We demonstrate a novel cross-sectional deformation, 
called the radial corrugation, 
of multi-walled carbon nanotubes (MWNTs) under hydrostatic pressure. 
Theoretical analyses based on the continuum elastic approximation
have revealed that MWNTs consisting of more than ten concentric walls undergo 
elastic deformations at critical pressure $p_c \simeq 1$ GPa, 
above which the circular shape of 
the cross section becomes radially corrugated. 
Various corrugation modes have been observed by tuning the innermost tube 
diameter and the number of constituent walls, which is a direct consequence of 
the core-shell structure of MWNTs.
}
}

\titlefigure[width=0.46\textwidth]{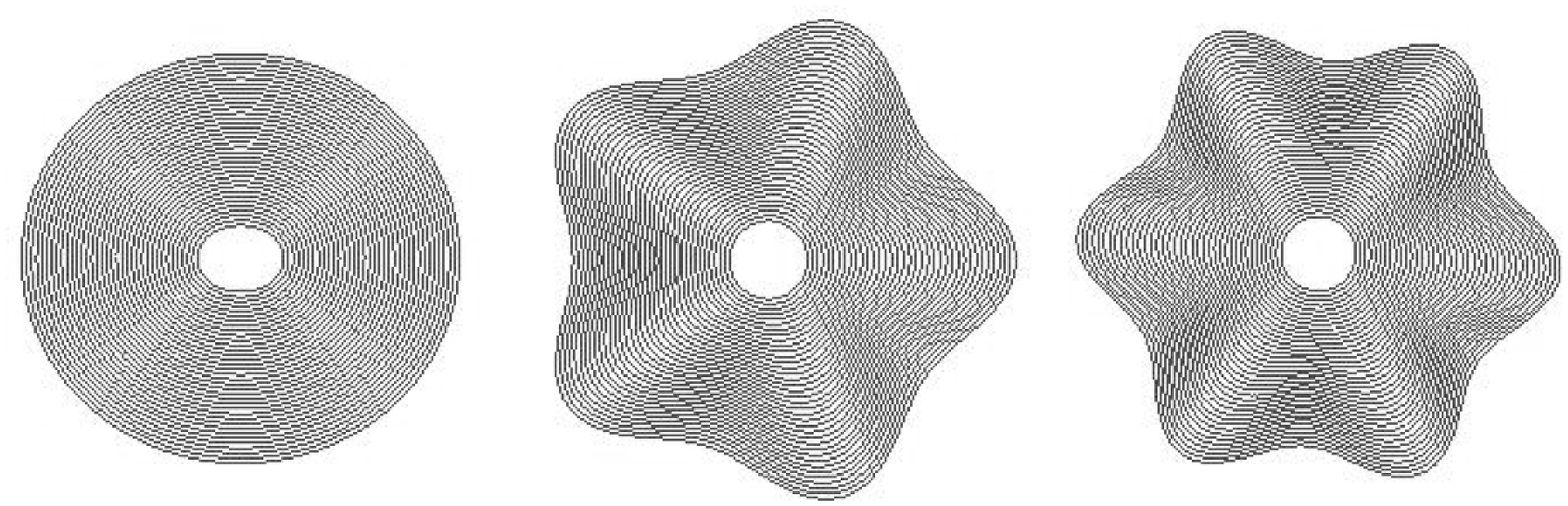}
\titlefigurecaption{Cross-sectional views of multi-walled carbon nanotube
under high hydrostatic pressure:
elliptic deformation with the mode index $n=2$ (left),
and radial corrugations with $n=5$ (center) and $n=6$ (right).
The index $n$ indicates the circumferential wave number of the deformed cross section.
}

\maketitle

\begin{figure*}[ttt]
\vspace*{-0.5cm}
\hspace*{-1cm}
\includegraphics*[width=10cm]{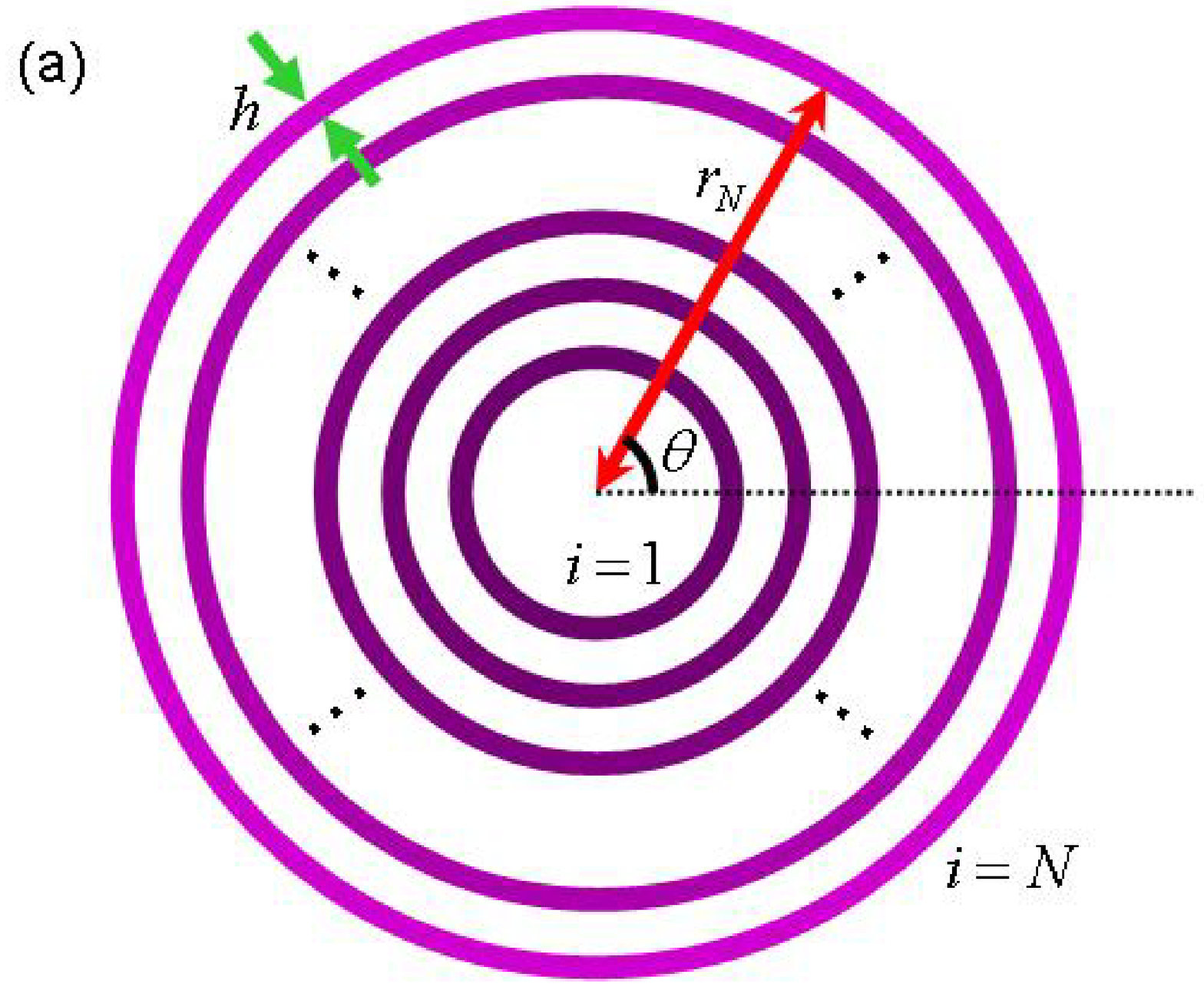}
\hspace*{-2cm}
\includegraphics*[width=10.5cm]{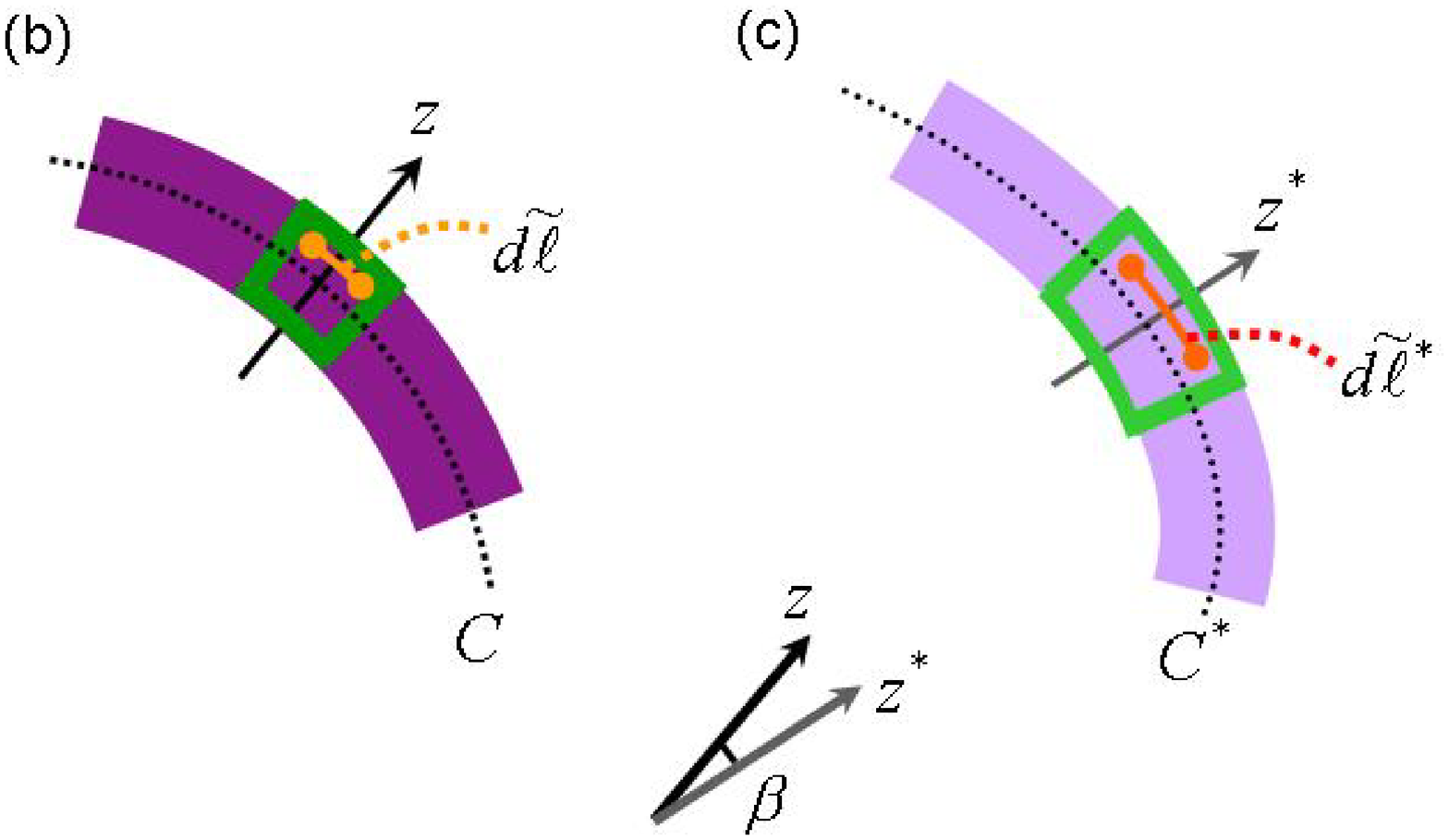}
\caption{Illustrations of geometric parameters of the continuum elastic-shell model.
(a) Sketch of cross section of MWNT consisting of $N$ cylindrical walls with thickness $h$.
Each wall is labeled by the index $i$.
(b) Enlarged view of a partial cross section of a {\it cylindrical} wall {\it before deformation}.
The circumferential line element $d\tilde{\ell}$
and the centroidal circle $C$ are depicted.
(c) Enlarged view of a partial cross section of a {\it deformed} wall under high pressure.
The line element is elongated in the circumferential direction,
and the normal to the centroidal curve rotates
with an angle $\beta$.}
\label{fig_01}
\end{figure*}

\section{Introduction}

An important mechanical feature of carbon nanotubes is their high flexibility in the radial direction. 
Radial stiffness of an isolated carbon nanotube is much less than 
axial stiffness \cite{Palaci2005}, which result in an elastic deformation of
the cross section on applying a hydrostatic pressure. 
Such a pressure-induced radial deformation yields significant changes in 
electronic \cite{Park1999,Mazzoni2000,DS_Tang2000,Gomez2006,Cai2006,Monteverde2006,Taira2007,Nishio2008}
and optical \cite{Venkateswaran1999,Deacon2006,Lebedkin2006,Longhurst2007}
properties, indicating the relevance of the deformation in carbon nanotube applications.
Thus far, many experimental and theoretical studies have been carried out on this issue
\cite{J_Tang2000,Peters2000,Sharma2001,Rols2001,Reich2002,Pantano2004,Elliott2004,Tangney2005,Gadagkar2006,Wang2006,Zhang2006,Hasegawa2006,Huang2006,Yang2006,Chrisofilos2007,Peng2008,Giusca2008,Wu2008,Jeong2008,Kuang2009,Lu2009}.
Most of them focused on single-walled nanotubes (SWNTs) and their bundles,
and revealed flattening and polygonalization in the cross section 
of SWNTs under pressures of the order a few GPa \cite{Venkateswaran1999,Sharma2001}.
Contrary to the intensive studies on SWNTs,
radial deformation of multiwalled nanotubes (MWNTs) still remains to be explored.
Intuitively, the multiple-shell structure of 
MWNTs is thought to enhance the radial stiffness of MWNTs.
However, when the number of concentric walls is much greater than unity, outside walls have 
large diameters so that external pressure may lead to a mechanical instability 
in the outer walls. 
This local instability implies a novel cross-sectional shape transition of MWNTs 
different from the cases of SWNTs.

This paper shows a new class of radial deformation, called {\it radial corrugation},
of MWNTs consisting of more than ten concentric walls.
In a corrugation mode, outside walls exhibit wavy structures along the circumference
as depicted in Titlefigure.
We demonstrate that various corrugation modes can
take place above critical hydrostatic pressure $p_c(N,D)$,
in which $p_c$ depends on the innermost tube diameter $D$
and the number of constituent walls $N$.
It should be emphasized that radial corrugation we have found
is a direct consequence of the core-shell structure of MWNTs,
and thus is differernt inherently from simple radial collapse observed in SWNTs.

\section{Continuum elastic-shell theory}

\subsection{Outline}

The stable cross-sectional shape of a MWNT under a hydrostatic pressure $p$
is evaluated by using
the continuum elastic theory for cylindrical shells \cite{Ru,Wang,Shen,He2005}.
The mechanical energy $U$ of a MWNT per unit axial length is written as
\begin{equation}
U = U\left[p, u_i(p,\theta), v_i(p,\theta) \right] = U_D + U_I + \Omega,
\label{eq_01}
\end{equation}
where $U_D$ is the deformation energy of all concentric walls,
$U_I$ is the interaction energy of all adjacent pairs of walls,
and $\Omega$ is the potential energy of the applied pressure.
All the three energy terms are functions of $p$ and 
the deformation amplitudes $u_i(p,\theta)$ and $v_i(p,\theta)$
that describe the radial and circumferential displacements, respectively,
of the $i$th wall.
See Eq.~(\ref{eq_aaa8}) below for the precise definitions of $u_i$ and $v_i$.

Our objective is the optimal displacements $u_i$ and $v_i$
that minimize the total energy $U$ under a given pressure $p$.
To this aim, we apply the variation method to $U$ with respect to $u_i$ and $v_i$,
and then obtain the stable cross-sectional shape of a MWNT under $p$ \cite{NTN}.
This strategy requires to derive explicit forms of $U_D$, $U_I$ and $\Omega$
as functions of $u_i$, $v_i$ and $p$,
which we shall resolve in the subsequent discussions.

\subsection{Strain-displacement relation}

We first consider the circumferential strain of a hollow cylindrical shell 
due to cross-sectional deformation.
Suppose a circumferential line element of length $d\tilde{\ell}$ lying within the cross section
of the shell with thickness $h$ (See Fig.~\ref{fig_01}).
The tilde $(\; \tilde{ } \;)$ attached to $\ell$ means that 
we consider the quantity at arbitrary point within the cross section.
We shall see later ({\it i.e.,} in Eq.~(\ref{eq_aaa8}) ) 
that the strain at arbitrary point
is determined approximately by the strain just on the controidal circle
denoted by $\cal{C}$ in Fig.~\ref{fig_01}.
This fact allows us to yield
a simplified relation between the strain and displacements of the shell
as given in Eqs.~(\ref{eq_A2_005}) and (\ref{eq_A2_006}).

The extentional strain $\tilde{\varepsilon}$
of the circumferential line element is defined by
\begin{equation}
\tilde{\varepsilon} = \frac{d\tilde{\ell}^* - d\tilde{\ell}}{d\tilde{\ell}}.
\label{eq_q04x}
\end{equation}
Here $d\tilde{\ell} = \tilde{r} d\theta$, and $d\tilde{\ell}^*$
is the length of the line element after deformation
(The asterisk symbolizes the quantity after deformation).
The coordinates $\tilde{x}^*$, $\tilde{y}^*$ of the element after deformation
is given by
\begin{eqnarray}
\tilde{x}^* (\theta) &=& 
\left[ \tilde{r} + \tilde{u}(\theta) \right] \cos \theta - \tilde{v}(\theta) \sin \theta, \nonumber \\
\tilde{y}^* (\theta) &=& 
\left[ \tilde{r} + \tilde{u}(\theta) \right] \sin \theta + \tilde{v}(\theta) \cos \theta,
\label{eq_app_1}
\end{eqnarray}
where $\tilde{u}$ and $\tilde{v}$ are components of the displacement vector
in the radial and circumferential directions, respectively.
It thus follows that
\begin{eqnarray}
& & \left( d\tilde{\ell}^* \right)^2 =
\left( d\tilde{x}^* \right)^2 + \left( d\tilde{y}^* \right)^2 \nonumber \\
&=&
\left[
\tilde{r}^2 + 2 \tilde{r} \left( \tilde{u}+\tilde{v} \right)
+ 
\left( \tilde{u}+\tilde{v} \right)^2 + \left( \tilde{u}-\tilde{v} \right)^2
\right] (d\theta)^2.
\label{eq_q05}
\end{eqnarray}

From Eq.~(\ref{eq_q04x}), we have 
$\tilde{\varepsilon}+1 = d\tilde{\ell}^*/d\tilde{\ell}$.
Hence, squaring the both sides and then rearranging the result give
\begin{equation}
\tilde{\varepsilon} + \frac12 \tilde{\varepsilon}^2
=
\frac12 \left[ \left( \frac{d\tilde{\ell}^*}{d\tilde{\ell}} \right)^2 - 1 \right].
\label{eq_q06}
\end{equation}
For $\tilde{\varepsilon} \ll 1$,
the term $\tilde{\varepsilon}^2$ can be omitted.
Hence, we have from Eqs.~(\ref{eq_q05}) and (\ref{eq_q06}) that
\begin{equation}
\tilde{\varepsilon} 
= \frac{\tilde{u}+\tilde{v}'}{\tilde{r}}
+ \frac12 \left( \frac{ \tilde{u}+\tilde{v}'}{\tilde{r}} \right)^2
+ \frac12 \left( \frac{ \tilde{u}'-\tilde{v}}{\tilde{r}} \right)^2,
\label{eq_q07}
\end{equation}
where $\tilde{u}' \equiv d\tilde{u}/d\theta$, {\it etc}.
The last term in Eq.~(\ref{eq_q07}) is
associated with the rotation of the line element
due to deformation.
The rotation angle $\beta$ consists of two parts:
i) a clockwise component 
$d\tilde{u}/d\ell = d\tilde{u}/(\tilde{r} d\theta)$ due to the spatial variation
of $\tilde{u}$ in the circumferential direction, and 
ii) a counterclockwise one $\tilde{v}/\tilde{r}$
due to the circumferential displacement of the element.
Combination of the two parts gives
\begin{equation}
\beta = \frac{\tilde{v} - \tilde{u}'}{\tilde{r}},
\end{equation}
which has a positive value in the counterclockwise sense.

The formula (\ref{eq_q07}) is valid for arbitrary large rotation $\beta$.
Particularly when $\tilde{\varepsilon}$ and $\beta$ are both sufficiently small (but finite),
we may neglect the second term in the right side in Eq.~(\ref{eq_q07}).
(Here we exclude the possibility that $|\tilde{u}|$ or $|\tilde{v}'|$
is of the order of $\tilde{r}$ or larger.)
We further assume that normals to the undeformed centroidal circle $\cal{C}$
remain straight, normal, and inextensional during the deformation (See Fig.~\ref{fig_01}).
As a result, $\tilde{u}$ and $\tilde{v}$ are expressed by
\begin{equation}
\tilde{u} = u \;\; \mbox{and} \;\; \tilde{v} = v + z \beta,
\label{eq_aaa8}
\end{equation}
where $u$ and $v$ denote the displacements of a point just on $\cal{C}$,
and $z$ is a radial coordinate measured from $\cal{C}$.
By substituting (\ref{eq_aaa8}) into (\ref{eq_q07}),
we attain the strain-displacement relation such as \cite{Sanders}
\begin{equation}
\tilde{\varepsilon}(z, \theta) = \varepsilon(\theta) + z \kappa(\theta),
\label{eq_A2_005}
\end{equation}
with the definitions:
\begin{equation}
\varepsilon = \frac{u + v'}{r} 
+
\frac12 \left( \frac{u' - v}{r} \right)^2
\;\; \mbox{and} \;\
\kappa = - \frac{{u}'' - {v}'}{r^2}.
\label{eq_A2_006}
\end{equation}
Here $r$ is the radius of the undeformed circle $\cal{C}$.
The results (\ref{eq_A2_005}) and (\ref{eq_A2_006})
state that the circumferential strain at arbitrary point in the cross section
is determined by the displacements $u(\theta)$ and $v(\theta)$
of a point just on the undeformed controidal circle $\cal{C}$.

\subsection{Deformation energy}

We are ready to derive the explicit form of the deformation energy $U_D$.
Suppose the $i$th cylindrical wall of a  long and thin circular tube with thickness $h$.
A surface element of the cross-sectional area of the wall is expressed by $r_i d\theta dz$.
The stiffness $k$ of the surface element for stretching along the circumferential direction
is given by 
\begin{equation}
k = \frac{E}{1-\nu^2},
\end{equation}
where $E$ and $\nu$ are Young's modulus and Poisson's ratio, respectively, of the wall.
Thus, the deformation energy $U_D^{(i)}$ of the $i$th wall per unit axial length
is written as
\begin{equation}
U_D^{(i)} = \frac{k r_i}{2} 
\int_{-h/2}^{h/2} \int_0^{2\pi} \tilde{\varepsilon}_i (z,\theta)^2 dz d\theta.
\label{eq_A2_004}
\end{equation}
From (\ref{eq_A2_004}) and (\ref{eq_A2_005}), we obtain
\begin{equation}
U_D^{(i)} = 
\frac{k h r_i}{2} \int_0^{2\pi} \varepsilon_i^2 d\theta
+
\frac{k h^3 r_i}{24} \int_0^{2\pi} \kappa_i^2 d\theta,
\label{eq_A2_007}
\end{equation}
which tells us the dependence of $U_D = \sum_{i=1}^N U_D^{(i)}$
on $u_i(\theta)$ and $v_i(\theta)$.

\begin{table}[b]
  \caption{Values of elasticity parameters used in this article.}
  \begin{tabular}[htbp]{@{}lll@{}}
    \hline
    Young's modulus & Poisson's ratio & Wall thickness \\
    \hline
    $E = 1$ [TPa]  & $\nu =0.27$ & $h= 0.34$ [nm] \\
    \hline
  \end{tabular}
  \label{onecolumntable}
\end{table}

\subsection{Inter-wall coupling energy}

Secondly, the explicit form of $U_I$ is described by
\begin{eqnarray}
U_I &=& \sum_{i=1}^{N-1} 
\frac{c_{i,i+1} r_i}{2}
\int_0^{2\pi} 
\left( u_i - u_{i+1} \right)^2 d\theta \nonumber \\
& & +
\sum_{i=2}^N
\frac{c_{i,i-1} r_i}{2}
\int_0^{2\pi} 
\left( u_i - u_{i-1} \right)^2 d\theta.
\end{eqnarray}
The vdW interaction coefficients $c_{ij}$ are functions of $r_i$ and $r_j$
and defined by \cite{He2005}
\begin{equation}
c_{ij} = 
- \left( 
\frac{1001 \pi \varepsilon \sigma^{12}}{3 a^4} F_{ij}^{13}
- \frac{1120 \pi \varepsilon \sigma^6}{9 a^4} F_{ij}^7
\right) r_j,
\label{eq_bb6}
\end{equation}
where $a$ denotes the chemical bond length between neignbouring carbon atoms 
within a layer ($a = 0.142$ nm),
and $\varepsilon$ and $\sigma$ are the parameters
that determine the vdW interaction between two layers 
($\varepsilon = 2.968$ meV and $\sigma = 0.3407$ nm) \cite{Saito}.
In Eq.~(\ref{eq_bb6}), we have set
\begin{equation}
F_{ij}^m = \left( r_i + r_j \right)^{-m}
\int_0^{\pi/2} \left( 1- K_{ij} \cos^2 \theta \right)^{-m/2} d\theta
\end{equation}
with $K_{ij} = 4 r_i r_j/ (r_i + r_j)^2$,
and the wall spacing $|r_i - r_{i\pm 1}| = 0.344 + 0.1 {\rm e}^{-D/2}$ nm
according to Ref.~\cite{Kiang}.

\begin{figure*}[ttt]%
\includegraphics*[width=0.47\textwidth]{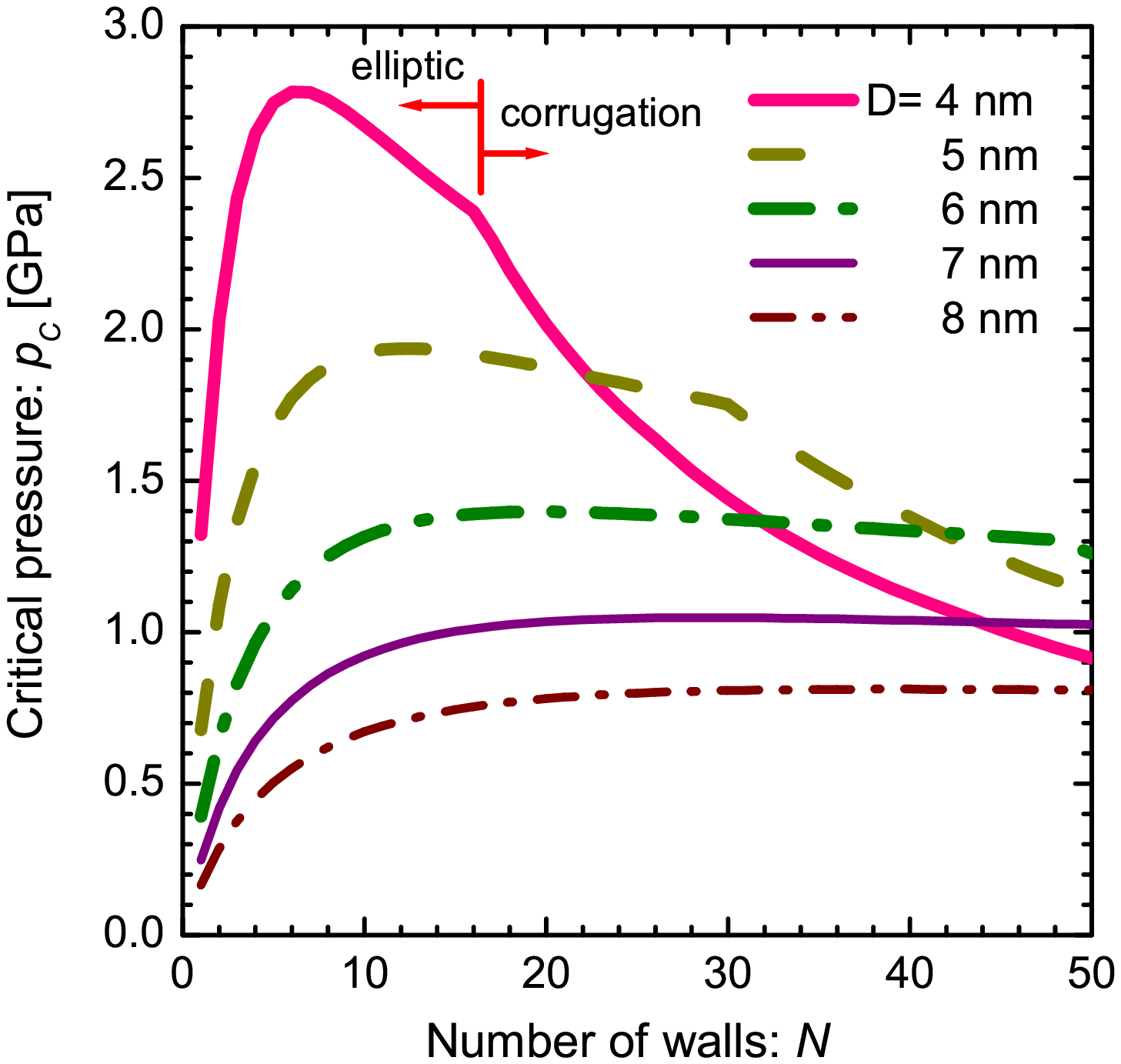}\hfill
\includegraphics*[width=0.47\textwidth]{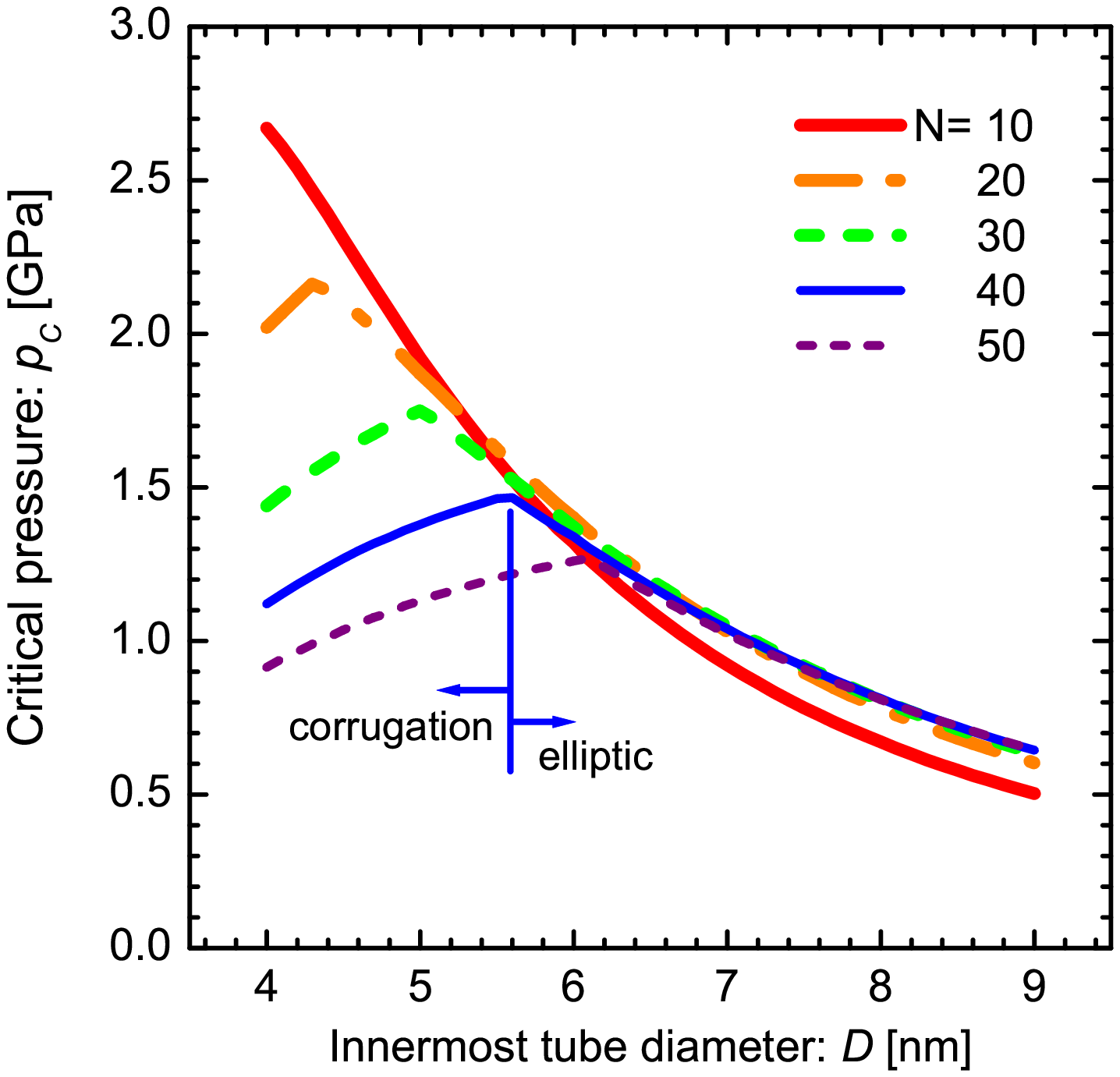}
\caption{\ \ Left: Wall-number dependence
of the critical pressure $p_c(N)$ for various innermost tube diameters $D$.
Each kink appearing at $N=N_0 =17$ for $D=4$ nm and $N_0=30$ for $D=5$ nm
separates the elliptic deformation phase $(N<N_0)$ and the radial corrugation phase $(N>N_0)$.
For $D\ge 6$ nm, only the elliptic deformation occurs independently of $N$.
\ Right: The $D$-dependence of $p_c(D)$ for various $Ns$.
Radial corrugation occurs at $D<D_0$, where $D_0$ indicates the position
of the kink appearing for $N\ge 20$.}
\label{fig_02}
\end{figure*}

\subsection{Pressure-induced energy}

We finally derive an explicit form of $\Omega$.
Since $\Omega$ is the negative of the work done by the external pressure
during cross-sectional deformation,
it is expressed as
\begin{equation}
\Omega = - p (\pi r_N^2 - S^*).
\label{eq_app_0}
\end{equation}
Here, $S^*$ is the area surrounded by the $N$th wall
after deformation (the sign of $p$ is assumed to be positive inward).
It is obtained by the line integral 
$S^* = \frac12 \oint_{\cal C} |\bm{r}_N^* \times d \bm{r}_N^*|$,
or equivalently
\begin{equation}
S^* = \frac12 \int_0^{2\pi} \left( x_N^* {y_N^*}' -y_N^* {x_N^*}' \right) d\theta.
\label{eq_app_2}
\end{equation}
From Eqs.~(\ref{eq_app_2}) and (\ref{eq_app_1})
as well as the periodicity relation $\int_0^{2\pi} {v_N}' d\theta = 0$,
we obtain
$$
\Omega = p \int_0^{2\pi} 
\left( r_N u_N + \frac{u_N^2 + v_N^2 - {u_N}' v_N + u_N {v_N}'}{2} \right) d\theta.
$$

\section{Critical pressure curve}

\subsection{Evaluating critical pressure $p_c$}

Our aim is to evaluate the critical pressure $p_c$
above which the circular cross section of a MWNT is elastically deformed
into non-circular one.
This is achieved by the following procedure \cite{NTN}.
First, we decompose the radial displacement terms as 
$u_i(p,\theta) = u_i^{(0)} (p) + \delta u_i(\theta)$,
where $u_i^{(0)} (p)$ indicates a uniform radial contraction at $p<p_c$
and $\delta u_i(\theta)$ describes a deformed, non-circular cross section
observed just above $p_c$.
Similarly, we can write
$v_i(p,\theta) = \delta v_i(\theta)$, since $v_i^{(0)}(p)\equiv 0$ at $p<p_c$.
Next, we apply the variation method to $U$ with respect to $u_i$ and $v_i$,
which results in a system of $2N$ linear differential equations
with respect to $\delta u_i$ and $\delta v_i$.
Then, substituting Fourier series expansions of $\delta u_i$ and $\delta v_i$
into the differential equations,
we obtain the matrix equation $\bm{C} \bm{u} = \bm{0}$.
Here, the vector $\bm{u}$ consists of 
the Fourier components of $\delta u_i$ and $\delta v_i$ 
with all possible $i$ and all Fourier components labeled by $n$,
and the matrix $\bm{C}$ involves one variable $p$
as well as elasticity parameters.
Finally, solving the secular equation ${\rm det}(\bm{C}) = 0$
with respect to $p$,
we obtain a sequence of discrete values of $p$;
among these $p$s, the minimum one serves as the critical pressure $p_c$
immediately above which the circular cross section of MWNTs
becomes radially deformed.

\subsection{Wall-number dependence}

Figure \ref{fig_02} shows numerical results of critical pressures $p_c$.
In the left-side figure,
$p_c$ is plotted as a function of 
the number of concentric walls $N$ with the innermost tube diameter $D$ being fixed.
For all $D$, $p_c$ increase with $N$ followed by a slow decay
or saturating constant value.
The increase in $p_c$ in the region of small $N$ is attributed to 
the enhancement of radial stiffness of the entire MWNT by encapsulation.
In contrast, the decay in $p_c$ indicates local instability in outside walls
that causes radial corrugation of MWNTs with $N \gg 1$.

\begin{figure*}[ttt]
\vspace*{-1cm}
\hspace*{-2cm}
\includegraphics*[width=0.7\textwidth]{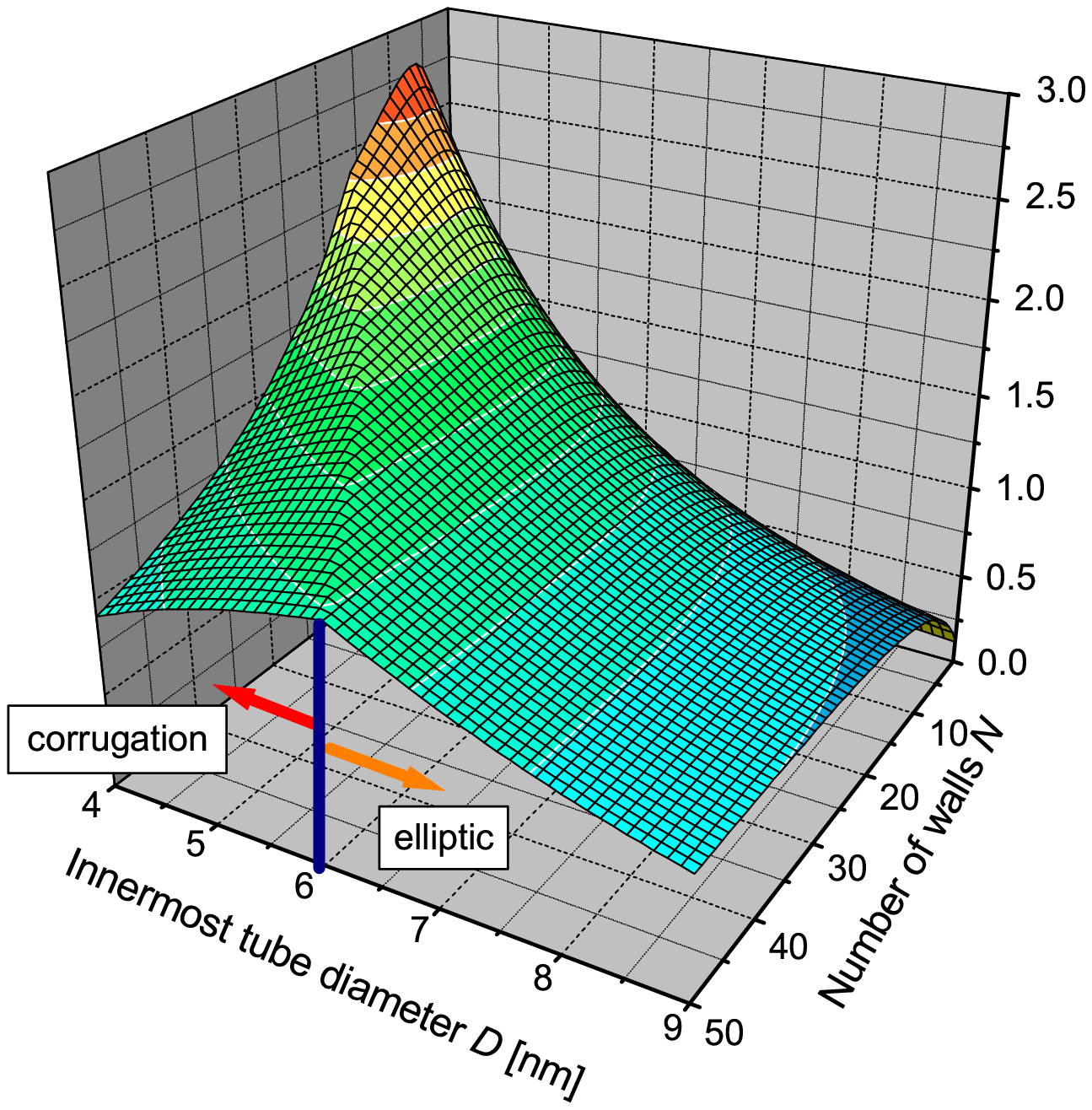}
\includegraphics*[width=0.4\textwidth]{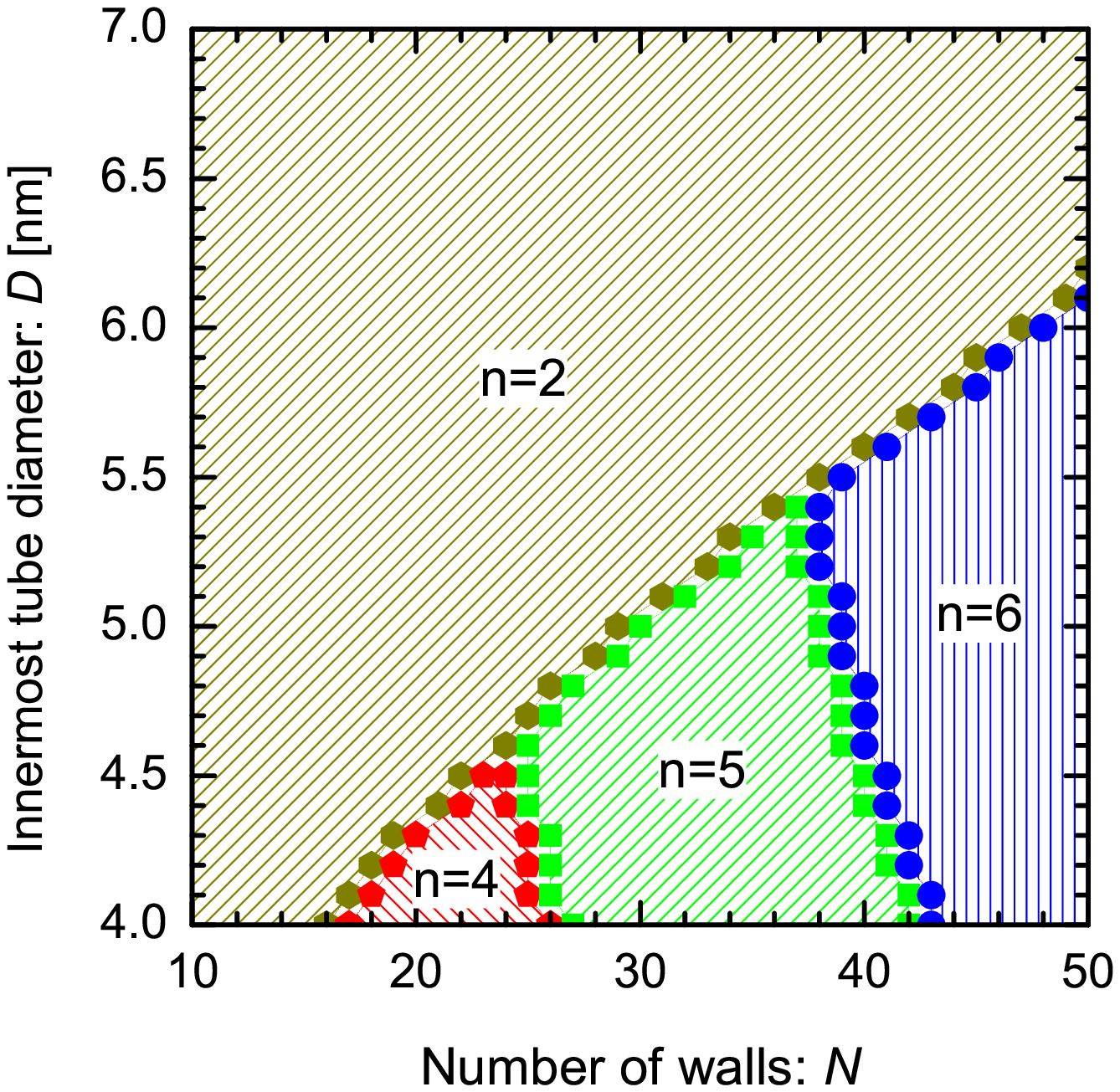}
\caption{\ \ Left: Three-dimensional plot of the critical pressure $p_c(N,D)$ 
in the $N$-$D$ plane.
A ridge line extending from the top to the bottom of the surface
corresponds to the phase-boundary kinks that separate the elliptic phase
from multiple corrugation phases.
Right: Phase diagram of the pressure-induced cross-sectional deformation
of MWNTs.}
\label{fig_03}
\end{figure*}

To be noteworthy is the occurrence of a kink in $p_c(N)$ for small $D$.
In the curve of $p_c(N)$ for $D=4$ nm, for instance, a kink appears at $N=N_0 =17$
to the right of which the value of $p_c$ decay monotonically.
The most striking is the fact that this kink determines the phase boundary 
between the elliptic deformation phase and the radial corrugation phases
(as indicated by red arrows in the left-sided figure of Fig.~2).
We have found that
$N$-walled nanotubes with $N>N_0$ exhibit some radial corrugation just above $p_c$,
while those with $N<N_0$ show only the elliptic deformation.
This senario also holds for the case of $D=5$ nm,
where the phase boundary kink occurs at $N_0 = 30$.
Which kind of corrugation mode ({\it i.e.}, what number of the mode index $n$)
takes place is dependent on the values of $N(>N_0)$ and $D$,
as clarified later in Fig.~\ref{fig_03}.

\subsection{Innermost-tube-diameter dependence}

The right-side figure in Fig.~\ref{fig_02}
illustrates the $D$-dependence of $p_c(D)$ for various $Ns$.
Similarly to the left-side figure,
a kink appears for each curve except for $N=10$.
This kink again serves as the phase boundary between the elliptic and corrugation phases;
if a kink occurs at $D_0$,
then it separates radial corrugation phases $(D<D_0)$
from the elliptic deformation phase $(D>D_0)$.

It is interesting that in the elliptic phase,
all the results (except for $N=10$) collapse onto a single curve.
This means that the number of walls $N$ is irrelevant 
in determining $p_c$ for MWNTs with large $D$.
In other words, encapsulation by increasing $N$ gives no contribution
to the radial stiffness of those MWNTs with large $D$.
This fact implies the availiability of 
simplified equations for estimating $p_c$ of MWNTs with large $D$;
establishing such simplified equations is currently in progress.

\section{Radial corrugation of MWNTs}

All the numerical results of $p_c$ are summarized 
into the three-dimensional plot $p_c(N,D)$;
see the left-side in Fig.~\ref{fig_03}.
This plot makes clear the optimal values of $N$ and $D$
that maximize (or minimize) the radial stiffness of MWNTs.
We also obverve a ridge line extending from the top to the bottom of the surface.
The ridge line corresponds to the phase-boundary kinks discussed in the previous section.
In fact, multiple corrugation modes take place in the region to the left of the ridge line,
as demonstrated in the phase diagram;
see the right-side of Fig.~\ref{fig_03}.
Interestingly, multiple corrugation modes are formed depending on the values of $N$ and $D$.
We see that smaller $D$ and larger $N$
favor corrugation modes, in which
larger $N$ yield higher corrugation modes with larger $n$.

\section{Concluding remarks}

We have demonstrated the presence of multiple radial corrugations
peculiar to MWNTs under hydrostatic pressures.
Theoretical investigations based on the continuum elastic theory have revealed that
MWNTs consisting of a large number
of concentric walls undergo elastic deformations at critical pressure $p_c \sim$ 1 GPa,
above which the cross-sectional circular shape becomes radially corrugated.
A phase diagram has been established
to obtain the requisite values of $N$ and $D$ for observing a desired corrugation mode.
It is hoped that our results should be verified by high-pressure experiments
on MWNTs as well as atomic-scale large-scale simulations \cite{Arias,Arroyo,Arroyo2}.

Another interesting subject is to explore the structural deformation effect
on the quantum-mechanical properties of $\pi$ electrons moving in the corrugated carbon walls.
It has been known \cite{curve1,curve2,curve3} 
that mobile electrons whose motion is confined to a two-dimensional curved thin layer
behave differently from those on a conventional flat plane;
this results from the occurrence of effective scalar- \cite{ShimaPRB,Ono} and vector- \cite{Taira}
potential energies induced by geometric curvature of the underlying layer.
Hence, quantum nature of electrons moving in the corrugated nanotube
will be strongly affected by geometric curvature of the walls,
which remains thus far unsettled.
Intensive studies on the issues mentioned above
will shed light on novel MWNT applications
based on cross-sectional deformation.

\begin{acknowledgement}
We acnowledge M.~Arroyo and R.~Dandoloff for stimulating discussions.
This study was supported by a Grant-in-Aid for Scientific Research from the MEXT, Japan.
One of the authors (H.S) is thankful for the financial support from 
Executive Office of Research Strategy in Hokkaido University.
A part of the numerical simulations were carried using
the facilities of the Supercomputer Center, ISSP, University of Tokyo. 
\end{acknowledgement}

\end{document}